\documentstyle[preprint,pre,aps,12pt]{revtex}
\begin{document}
\title{On the simplest (2+1) dimensional integrable spin systems and their
equivalent nonlinear Schr\"odinger equations}
\author{R. Myrzakulov$^{{a,b}*}$, S. Vijayalakshmi$^{c}$, 
R.N. Syzdykova$^{b}$ and M. Lakshmanan$^{c{\dag}}$}
\address {$^{a}$ Physical Technical Institute, National Academy of Sciences,
Alma-Ata-480 082, Kazakstan\\ 
$^{b}$ Center for Nonlinear Problems, PO Box 30, 480035 Alma-Ata-35,
Kazakstan \\
$^{c}$  Centre for Nonlinear Dynamics,
Department of Physics, Bharathidasan University, Tiruchirapalli 620 024,
India}
\maketitle
\begin{abstract}
Using a moving space curve formalism, geometrical as well as gauge equivalence 
between a (2+1) dimensional spin equation (M-I equation) and the (2+1) dimensional 
nonlinear Schr\"odinger equation (NLSE) originally discovered by Calogero, discussed 
then by Zakharov and recently rederived by Strachan, have been estabilished. A 
compatible set of three linear equations are obtained and integrals of motion are 
discussed. Through stereographic projection, the M-I equation has been 
bilinearized and different types of solutions such as line and curved solitons,
breaking solitons, induced dromions, and domain wall type solutions are
presented. Breaking soliton solutions of (2+1) dimensional NLSE have also been
reported. Generalizations of the above spin equation are discussed.
\end{abstract}
\section{INTRODUCTION}
The two dimensional sigma-models with nontrivial topological structures are
known to play a useful role in modern field theory. This set of models
constitute a laboratory for studying two dimensional analogues of elementary
particles within the framework of classical field theory[1]. In this context,
important case studies concern with the existence of localized coherent structures
(dromions, lumps, etc.) and other types of soliton-like solutions of some
classical nonlinear field models such as, for example, the Ishimori and the
other (2+1) dimensional spin systems[1-3].

Generally speaking, in (2+1) dimensions we have a number of remarkable properties, 
which may not exist in (1+1) dimensionsal counterparts. For example,

$1^{\circ}$. These equations possess the so called localized coherent structures 
(such as dromions, lumps and so on), which may undergo both elastic and inelastic 
scattering depending upon the initial conditions.

$2^{\circ}$.The corresponding spectral parameter (eigenvalue) of the associated 
linear problem (the Lax representation) can be dependent on the $t$ (time) and 
$y$ (space) variables and satisfy even nonlinear equations (see, for instance,
[3-4]). As a consequence, the original soliton equation can have breaking 
solutions[4-6]. Besides, in this case, for finding solutions one even needs to use 
non-isospectral Inverse Scattering Transform(IST)[5].

$3^{\circ}$. Each integrable (1+1) dimensional equation may admit several 
integrable (and nonintegrable) extensions[2,3] in (2+1) dimensions. For example, 
the KdV equation has integrable extensions such as KP, NNV and breaking soliton 
equations[6].

$4^{\circ}$. Regarding the subject of the present paper, many of the (2+1)
dimensional spin equations possess the topological invariant-with the so called
topological charge

$$Q={1 \over {4\pi }} \int dx dy \vec S \cdot \vec S_x \wedge \vec S_y,
\eqno(1)$$
and their solutions are classified by the integer values of $Q_N$, $ N=0,\pm 1,
\pm 2,$... Here, $\vec S = (S_1, S_2, S_3) $ and $\mid \vec S \mid^2=S_1^2+
S_2^2+S_3^2=1$ and $\vec S_x = \left( {\partial \vec S\over {\partial x}}\right)$,
$\vec S_y = \left( {\partial \vec S\over {\partial y}}\right)$. This property 
can be realised, for example, in the (2+1) dimensional analogues of the well 
known (1+1) dimensional Heisenberg ferromagnet model (or the (1+1) dimensional 
isotropic Landau - Lifshitz equation (LLE))[7-9]

$$\vec S_{t} = \vec S \wedge \vec S_{xx}.   \eqno(2)$$

The Heisenberg ferromagnetic spin equation (2) possesses many useful (2+1)
dimensional extensions: Some of them are as follows.

a) The Ishimori equation[10]

$$\vec S_{t} = \vec S \wedge ( \vec S_{xx} +\epsilon^2 \vec S_{yy})
+ u_{x} \vec S_{y} + u_{y} \vec S_{x},    \eqno(3a)$$

$$u_{xx} -\epsilon^2 u_{yy} = -2\epsilon^2 \vec S \cdot (\vec S_{x}
\wedge \vec S_{y}). \eqno(3b)$$

b) The M - I equation[5]

$$\vec S_{t} = (\vec S \wedge \vec S_{y} + u \vec S )_{x},  \eqno(4a)$$

$$u_{x} = - \vec S \cdot (\vec S_{x} \wedge \vec S_{y}).    \eqno(4b)$$

c) The (2+1) dimensional isotropic LLE[11]

$$\vec S_{t} = \vec S \wedge (\vec S_{xx} + \vec S_{yy}).   \eqno(5)$$
Here $u$ is a scalar function and $\epsilon^2=\pm 1$. It turns out that eqs.
(3) and (4) are integrable, while Eq. (5) is apparently nonintegrable. All the
three Eqs. (3) - (5) describe the nonlinear dynamics of the classical spin
systems in the plane and in the (1+1) dimensional case reduce to one and the
same Eq. (2). Properties of Eqs. (3) are relatively well studied (see, for
example, [3,10]). In this paper we wish to concentrate on studying the M-I
equation(4). In particular, we wish to identify the geometrically equivalent
(2+1) dimensional NLSE for (4) and study the nature of the nonlinear
excitations admitted by the spin system.

The paper is organized as follows. In Sec.II we briefly review some necessary
informations about (4) relevant to this paper. The geometrical and gauge
equivalent counterpart of Eq. (4) is constructed in Sec.III and Sec.IV,
respectively. Integrals of motion are discussed in Sec.V. In Sec.VI the Hirota 
bilinear form of Eq. (4) is derived. The solitons (line and curved), domain walls 
and dromion-like solutions as well as their breaking analogues are obtained in 
Sec.VII and breaking solitons and dromions for its equivalent counterpart are 
obtained in Sec.VIII. In Sec.IX we comment on the possible further extensions 
and we conclude in Sec.X.  
\section{THE M-I EQUATION }

The Lax representation of Eq. (4) can be shown to take the form[5]

$$\phi_{1x} = U_{1}\phi_1,   \eqno(6a)$$

$$\phi_{1t} = V_{1} \phi_1 + \lambda \phi_{1y},  \eqno(6b)$$
with

$$U_{1} = \frac{i\lambda}{2} S ,                 \eqno(7a)$$

$$V_{1} = \frac{\lambda }{4} ([S,S_{y}] + 2iuS).   \eqno(7b)$$
Here

$$S=\pmatrix{
S_3 & rS^-\cr
rS^+ & -S_3},         \eqno(8)$$
$S^{\pm} = S_{1} \pm iS_{2}$ and $\lambda $ is the eigenvalue parameter, which 
satisfies the following nonlinear equation

$$\lambda_{t}=\lambda \lambda_{y}. \eqno(9)$$
Hence, using the compatibility of Eqs. (6a) and (6b), we get

$$iS_t=\left( [S,S_y]+2iuS\right)_x , \eqno(10a) $$

$$u_x=-{1\over 2i} tr(S S_x S_y),   \eqno(10b)$$
where tr denotes the trace of the matrix. This system is obviously the
matrix form of Eq. (4).

Thus, for solving equation(4), we must use the non-isospectral extension of IST
[5]. As a consequence, Eq. (4) admits besides the usual solutions, corresponding 
to constant solution of Eq. (9), breaking analogues related to the $t$ and $y$ 
dependence of $\lambda $. Note that Eq. (9) itself is the compatibility condition 
of the following system of the linear equations,

$$f_x={i\lambda \over 2}f, \eqno(11a)$$

$$f_t=\lambda f_y. \eqno(11b)$$

Another useful form of Eq. (4), can be obtained by using the complex (stereographic) 
variable $\omega (x,y,t)$ defined through the relations

$$S^{+} = S_{1}+iS_{2}=\frac {2\omega}{1+\mid \omega \mid^{2}},
\,\,\,\,\, S_{3}=\frac{1-\mid \omega \mid^{2}}{1+\mid \omega \mid^{2}}.
\eqno(12)$$
In this case, Eq. (4) takes the form

$$i(\omega_{t} - u\omega_{x}) + \omega_{xy} -
\frac{2\omega^* \omega_{x}\omega_{y}}{1 + \mid\omega\mid^{2}}
= 0, \eqno(13a)$$

$$u_{x} + \frac{2i(\omega_{x}\omega^{*}_{y} - \omega^{*}_{x}\omega_{y})}
{(1 + \mid \omega \mid^{2})^{2}} = 0.     \eqno(13b)$$
We will use this form also frequently in our following analysis.
\section{L - EQUIVALENT COUNTERPART}

It is well known that in (1+1) dimensions there exists geometrical equivalence
between spin systems and nonlinear Schr\"odinger type equations[7,12], which
in ref.[5] was called the Lakshmanan eqivalence or shortly the L-equivalence
(see also [12-14]). In refs.[5,13-15] a (2+1) dimensional generalization of 
the L-equivalence was presented. In this section we find the L-equivalent 
counterpart of Eq. (4). For this purpose, we will extend the geometrical method 
applicable to (1+1) dimensional systems suitably to the (2+1) dimensional case. 
We now associate a moving space curve parametrised by the arclength $x$, and 
endowed with an additional coordinate $y$, with the spin system[12,16]. Then 
the Serret-Frenet equation associated with the curve has the form

$$ \vec e_{ix} = \vec D \wedge \vec e_i,   \eqno (14) $$
where

$$ \vec D = \tau \vec e_1 + \kappa \vec e_3    \eqno(15)$$
and $ \vec e_i$'s, $i = 1,2,3,$  form the orthogonal trihedral.

Mapping the spin variable on the unit tangent vector

$$ \vec S(x,y,t) = \vec e_1 , \eqno(16) $$
the curvature and the torsion are given by

$$ \kappa (x,y,t) = ( \vec S_x^2)^{1\over2},  \eqno(17a)$$

$$\tau(x,y,t) = \kappa ^{-2} \vec S \cdot (\vec S_{x} \wedge \vec S_{xx}).
\eqno(17b)$$
Due to the  orthonormality nature of the trihedral, $\vec e_{it}. \vec e_i = 0$, 
$\vec e_{iy}. \vec e_i = 0$,  $i,j = 1,2,3$ and using the compatibility condition
$ \vec e_{ixy} = \vec e_{iyx}$, we find the equation for the $y$-part

$$ \vec e_{iy} = \vec {\gamma} \wedge \vec e_i, \eqno(18)$$
where $\vec {\gamma} =(\gamma_1,  \gamma_2, \gamma_3) $ and

$$ \gamma_1 = u+\partial_x^{-1} \tau_y ,  \eqno(19a)$$

$$ \gamma _2 = {u_x \over \kappa },     \eqno(19b)$$

$$\gamma_3= \partial_x^{-1}\left( \kappa _y-{\tau u_x \over \kappa}
\right ).   \eqno(19c)$$

Alternatively the trihedral $\vec e_i(x,y,t)$, $i=1,2,3$ could be thought of as 
defining a suitable surface in $E^{3}$, (so that Eqs. (14) and (18) represent 
the Gauss - Weingarten equations in orthogonal coordinates and that the 
compatibility condition $\vec e_{ixy}=\vec e_{iyx}$  gives rise to the Codazzi
-Mainardi equations), which is then set in motion. Now, from Eq. (4) and using 
Eqs. (14) and (18), we can easily find the time evolution of the trihedral as

$$ \vec e_{it} = \vec {\Omega} \wedge \vec e_i, \eqno(20)$$
with

$$ \vec \Omega = (\omega_1, \omega_2, \omega_3)
          = \left( {\kappa_{xy} \over \kappa }-\tau \partial_x^{-1} \tau_y, -\kappa_y,
          -\kappa \partial_x^{-1} \tau_y \right). \eqno(21)  $$
Ultimately, the compatibility condition  $\vec e_{ixt} =\vec e_{itx}$, which is 
also consistent with the relation $\vec e_{iyt} =\vec e_{ity}$, $i=1,2,3$ yields 
the following evolution equations for the curvature and torsion,

$$ \kappa_t= -(\kappa \tau)_y-\kappa_x \partial_x^{-1} \tau_y  ,\eqno(22a)$$

$$ \tau_t= \left[ {\kappa_{xy}\over {\kappa}}- \tau \partial_x^{-1} \tau_y
\right]_x + \kappa \kappa_y .  \eqno(22b)$$
On making the complex transformation[7]

$$ \psi (x,y,t) ={\kappa(x,y,t)\over 2} \exp \left [-i\int_ {-\infty }^{x}
\tau(x',y, t) dx'\right ], \eqno(23)$$
the set of equations (22) reduces to the following (2+1) dimensional NLSE

$$ i\psi_t = \psi_{xy}+ r^2V \psi, \eqno(24a)$$

$$ V_x = 2\partial_y{\mid \psi \mid }^2.     \eqno(24b)$$
Here, $r^2 = +1$, that is, we have the attractive type NLSE (The case $r^2
= -1$ corresponds to the repulsive case). Eq. (24) belongs to the class of equations
discovered by Calogero[17] and then discussed by Zakharov[18] and recently 
rederived by Strachan (for $r^2 = +1$)[19]. Its Painlev\'e property and some 
exact solutions were also obtained [20]. $N$-soliton solutions of Eq. (24) for 
both the cases ($r^2=\pm 1$) can be found in ref.[21]. Thus, we have proved that 
Eq. (24) is equivalent to Eq. (4) in the geometrical sense.
\subsection{LINEARIZATION}

Introducing now the complex variable corresponding to an orthogonal rotation

$$z_l={e_{2l}+ie_{3l}\over {1-e_{1l}}},\,\,\,\,e_{1l}^2+e_{2l}^2+e_{3l}^2=1,\,\,
\,\,\,l=1,2,3 $$
the spatial and temporal evolution of the trihedral (eqs. 14, 18 and 20) can
be rewritten as a set of the following three Riccati equations:

$$z_{lx}=-i\tau z_l+{\kappa \over 2}\left[ 1+z_l^2\right],\eqno(25a)$$

$$z_{ly}=-i\gamma_1 z_l+{1\over 2}\left[ \gamma_3+i\gamma_2\right] z_l^2+
{1\over 2}\left[ \gamma_3-i\gamma_2\right],\eqno(25b)$$

$$z_{lt}=-i\omega_1 z_l+{1\over 2}\left[ \omega_3+i\omega_2\right] z_l^2+
{1\over 2}\left[ \omega_3-i\omega_2\right].\eqno(25c)$$
It is easy to check that Eq. (25a) is equivalent to the Serret-Frenet equations
(14), (25b) is equivalent to the $y$-variation of the trihedral Eq. (18), while
the temporal evolution (25c) is equivalent to (20).

Further introducing the transformation

$$z_l={v_2\over{v_1}},\eqno(26)$$
Eq. (25) can be written as a system of three coupled two component first order
equations,

$$\pmatrix {
v_{1x} \cr
v_{2x}
}  = \pmatrix{
{i\tau \over 2} & {-\kappa \over 2} \cr
{\kappa \over 2} & {-i\tau \over 2}
}
\pmatrix{
v_1 \cr
v_2
}, \eqno(27a) $$

$$\pmatrix {
v_{1y} \cr
v_{2y}
}  = \pmatrix{
{i\gamma_1\over 2} & {{-1\over 2}(\gamma_3+i\gamma_2)} \cr
{{1\over 2}(\gamma_3-i\gamma_2)} & {-i\gamma_1\over 2}
}
\pmatrix{
v_1 \cr
v_2
}, \eqno(27b) $$

$$\pmatrix {
v_{1t} \cr
v_{2t}
}  = \pmatrix{
{i\omega_1\over 2} & {{-1\over 2}(\omega_3+i\omega_2)} \cr
{{1\over 2}(\omega_3-i\omega_2)} & {-i\omega_1\over 2}
}
\pmatrix{
v_1 \cr
v_2
}. \eqno(27c) $$
Once again one can check that the compatibility of the three sets of
equations (27) gives rise to the evolution equation (22).
\section{GAUGE EQUIVALENT COUNTERPART}

Next, it is also of interest to note that Eqs. (4) and (12) are
gauge equivalent to each other in the sense of Zakharov and Takhtajan[22].
To obtain the gauge equivalent counterpart of Eq. (4), in the usual way
we consider the following gauge transformation

$$\phi_1 = g^{-1} \phi_2 , \eqno(28)$$
where $g(x,y,t)$ and $\phi_2(x,y,t,\lambda )$ are arbitrary $(2\times 2)$ matrix
functions of the type defined on a compact manifold $S^2 = SU(2)/U(1)$. 
Substituting Eq. (28) into Eq. (6), after some algebra  we get the following 
system of linear equations for $\phi_2$,

$$ \phi_{2x}= U_2 \phi_2 , \eqno(29a)$$

$$\phi_{2t} = V_2 \phi_2 + \lambda\phi_{2y} \eqno(29b)$$
with

$$ U_2= {i\lambda\over2} \sigma_3+G,\,\,\,\,G= \pmatrix{
 0 & \phi \cr
 -r^2\phi^* & 0
 } ,\,\,\,\,r^2=+1,\eqno(30a)$$

$$V_2= -i\sigma_3 \left({VI\over2}+G_y\right),\,\,\,\, I=diag(1,1), \eqno(30b)$$

$$ V= 2 \partial_x^{-1} \partial_y\left(\mid \psi\mid^2\right). \eqno(30c)$$
The compatibility condition of Eq. (29) along with (9) becomes (24), that is, Eq. (4)
and Eq. (24) are gauge equivalent to each other. The above transformation is in
fact reversible and we can similarly prove that Eq. (24) is gauge equivalent to
Eq. (4). It is also of interest to note that the set of linear Eqs. (27) can be recast
in the form (29) after suitable transformations.

Next, we present some important formulae which are just consequences of the
geometrical/gauge equivalence of Eqs. (4) and (24). We have

$$ tr(S_x^2) = 8 \mid \psi \mid ^2= 2 \vec S_x^2. \eqno(31a) $$
In a similar manner we find that

$$ -2i \vec S \cdot (\vec S_x \wedge \vec S_{xx})=tr(SS_xS_{xx})= 4(\psi^* \psi
_x-\psi \psi^*_x ). \eqno(31b) $$
These relations are obviously equivalent to Eq. (23). One  notes that these are 
of the same form as in the case of (1+1) dimensional Heisenberg spin chain [7,8]. 
\section{INTEGRALS OF MOTION}

The spin Eq. (4) allows an infinite number of integrals of motion as a consequence
of integrability. These integrals can be consructed using for example the Lax 
representation(6). However some of the integrals can be constructed using that
L-equivalence property. Now from (22a), it follows that

$$(\kappa^{2})_{t} = [-\kappa^{2}\partial^{-1}_{x}\tau_{y}]_{x} +
[-k^{2}\tau]_{y}.\eqno(32a)$$
Hence we get the first integral

$$K_{1} =\int \kappa^{2} dxdy .\eqno(32b)$$
Similarly (22b) leads to

$$K_{2} = \int \kappa^2\tau dxdy .\eqno(33)$$
In terms of the spin vector the corresponding  two conservation laws are

$$\left( \vec S^2_x\right) _{t}+ \partial _x{\left[\vec S_x^2
\partial _x^{-1}\left( \vec S\cdot \vec S_x\wedge \vec S _{xx}\over {\vec S_x^2} 
\right)_y\right]} + \partial_y \left[ \vec S \cdot \vec S_x\wedge 
\vec S_{xx}\right]=0,\eqno(34)$$

$$ \left[ \vec {S}\cdot \vec S_x \wedge \vec S_{xx} \right]_t +
\partial _x \left[ {(\vec S_x^2)_x (\vec S_x^2)_y \over {4\vec S_x^2}} + \vec S 
\cdot \vec S_x \wedge \vec S_{xx} \partial
_x^{-1} \left( {\vec {S}\cdot \vec S_x \wedge \vec S_{xx}\over {\vec S_x^2}} 
\right)_y\right]$$
$$ + \partial_y \left\{ {(\vec S_x^2)^2_x\over {4\vec S_x^2}} + {{ \left( \vec 
{S}\cdot \vec S_x \wedge \vec S_{xx} \right)^2} \over {\vec S^2_x}} -
{\left( \vec S_x^2 \right)_{xx}\over 2}-{\vec S_x^4 \over 4}\right\} = 0 .
\eqno (35)$$
Note that these integrals have the same forms as in the (1+1) dimensional case. 
More interesting integrals of purely (2+1) dimensional nature can be obtained 
from the condition

$$\vec e_{jxy} = \vec e_{jyx}.\eqno(36)$$
Making use of the various relations in Sec.III and after some algebra we have

$$(-\kappa \gamma_2)_t = (-\gamma_{1t})_x + (\tau_t)_y,\eqno(37) $$

$$(-\tau \gamma_2)_t = (\gamma_{3t})_x + (-\kappa_t)_y. \eqno(38) $$
These equations give two integrals

$$K_3 = \int (-\kappa \gamma_2) dxdy, \eqno(39) $$
and 

$$K_4 = \int (-\tau \gamma_2) dxdy . \eqno(40)$$

In terms of the spin vector $\vec S $ these integrals of motion take the forms

$$K_1 = \int (\vec S_x^2 )dxdy, \eqno(41)$$

$$K_2 =  \int \vec S \cdot \vec S_{x} \wedge \vec S_{xx}
dxdy, \eqno(42)$$

$$K_3 = \int \vec S \cdot (\vec S_{x} \wedge \vec S_{y} )dxdy \eqno(43)$$
and

$$K_4 = \int \frac {[\vec S \cdot (\vec S_{x}\wedge \vec S_{y})]
[\vec S \cdot (\vec S_{x} \wedge \vec S_{xx})]}
{(\vec S_x^2){\frac{3}{2}}}dxdy .\eqno(44)$$
Note that $K_3$ is the topological charge given by Eq. (1) to within a constant. 
One can proceed to find the other integrals of motion using the eigenvalue 
problem(29).
\section{HIROTA BILINEAR FORM}

In order to solve the Cauchy initial value problem of Eq. (4),
it will be of interest to investigate the system in the framework of the
inverse spectral transform method, for example, by the d-bar dressing
method[2,3]. However, for our present purpose we concentrate on exact analytic
solutions of Eq. (4). In doing so, it is convenient to rewrite Eq. (4) or its
equivalent stereographic form (13) in the Hirota bilinear form. On writing

$$\omega={g\over f}, \eqno(45)$$
Eq. (4) or Eq. (13) becomes

$$ (iD_t-D_xD_y) (f^*\circ g)=0, \eqno(46a)$$

$$ (iD_t-D_xD_y) (f^*\circ f-g^*\circ g)=0, \eqno(46b)$$

$$D_x(f^*\circ f+g^*\circ g)=0,\eqno(46c)$$
while the potential $u$ takes the form

$$u(x,y,t)=-{iD_y(f^*\circ f+g^*\circ g)\over {f^*\circ f+g^*\circ g}},
\eqno(46d)$$
where $g$ and $f$ are complex valued functions.
Here $D_{x}$  is the Hirota bilinear operator, defined by

$$D^{k}_{x} D^{m}_{y} D^{n}_{t} (f \circ g) = (\partial_x-\partial_
{x\prime })^k(\partial_y-\partial_{y\prime })^m (\partial_t-\partial_{t\prime })^n
f(x,y,t)g(x,y,t) \|_{x=x\prime, y=y\prime, t=t\prime} \eqno(47)$$
Using the above definition of the $D$-operator, we get from (31d) that

$$u_x = - 2i\left[ D_y (f \circ g) D_x(f^*\circ g^*)- c.c \right].\eqno(48a)$$
In terms of $g$ and $f$, the spin field takes the form

$$S^{+} = \frac{2f^{*}g}{\mid f\mid ^{2} + \mid g \mid ^{2}},\,\,\,\,\,
S_{3}=\frac{\mid f\mid^{2}-\mid g\mid^{2}}{\mid f\mid^{2}+\mid g\mid^{2}}.
\eqno(48b)$$
Eq. (46) represents the starting point to obtain interesting classes of
solutions for the spin system (4).The construction of the solutions is standard.
One expands the functions $g$ and $f$ as a series

$$g = \epsilon g_{1} + \epsilon^{3} g_{3} + \epsilon^{5}g_{5} +
\cdot \cdot \cdot \cdot \cdot, \eqno(49a)$$

$$f=1+\epsilon^2 f_2+\epsilon^4 f_4+\epsilon^6 f_6+ ..... . \eqno(49b)$$

Substituting these expansions into (46 a,b,c) and equating the coefficients
of $\epsilon $, one obtains the following system of equations from (46a):

$$\epsilon^1: ig_{1t}+g_{1xy}=0 ,\eqno(50a)$$

$$\epsilon^3: \left[ i\partial_t+\partial_x \partial_y\right] g_3=\left[iD_t
-D_xD_y\right] (f^*_2.g_1),\eqno(50b)$$
$$  \cdot                 \cdot $$
$$  \cdot                 \cdot $$
$$  \cdot                 \cdot $$
$$\epsilon^{2n+1}: \left[ i\partial_t+\partial_x \partial_y\right] g_{2n+1}
=\sum_{k+m=n} \left[ iD_t-D_xD_y\right] (f^*_{2k}.g_{2m+1}),\eqno(50c)$$
and from (46b):

$$\epsilon^2: i\partial_t(f_2^*-f_2)-\partial_x \partial_y(f_2^*+f_2)=
\left[iD_t-D_xD_y\right] (g_1^*.g_1),\eqno(51a)$$

$$\epsilon^4: i\partial_t(f_4^*-f_4)-\partial_x \partial_y (f_4^*+f_4)=
\left[iD_t-D_xD_y\right] (g_1^*.g_3+g_3^*.g_1-f_2^*.f_2),\eqno(51b)$$
$$  \cdot                 \cdot $$
$$  \cdot                 \cdot $$
$$  \cdot                 \cdot $$

$$\epsilon^{2n}: i\partial_t(f_{2n}^*-f_{2n})-\partial_x \partial_y (f_{2n}^*
+f_{2n})=(iD_t-D_xD_y)\left( \sum_{n_1+n_2=n-1}g_{2n_1+1}^*.g_{2n_2+1}\right)$$

$$-(iD_t-D_xD_y)\left(\sum_{m_1+m_2=n}f_{2m_1}^*.f_{2m_2}\right).\eqno(51c)$$
Further from (46c), we have the following:

$$\epsilon^2: \partial_x (f_2^*-f_2)=-D_x(g_1^*.g_1),\eqno(52a)$$

$$\epsilon^4: \partial_x (f_4^*-f_4)=-D_x(g_1^*.g_3+g_3^*.g_1+f_2^*.f_2),
\eqno(52b)$$
$$  \cdot                 \cdot $$
$$  \cdot                 \cdot $$
$$  \cdot                 \cdot $$

$$\epsilon^{2n}: \partial_x (f_{2n}^*-f_{2n})= -
D_x\left[ \sum_{n_1+n_2=n-1}(g_{2n_1+1}^*.g_{2n_(2+1)}+
\sum_{n_1+n_2=n}f_{2n_1}^*.f_{2n_2})\right].\eqno(52c)$$
Solving recursively the above equations, we obtain many interesting classes of
solutions to Eq. (4).
\section{SOLUTIONS OF THE SPIN SYSTEM}

Using the results of the previous section, we are in a position to construct
many exact solutions such as solitons, domain walls, breaking solitons and
induced dromions of Eq. (4). To obtain such solutions, we can use Eqs. (46) as
the starting point.
\subsection{The 1-line soliton and curved soliton solutions}

On solving Eq. (50a), we obtain

$$ g_1 = \sum_{j=1}^N \exp {\chi_j},\,\,\,\,\chi_j = a_jx + b_j(y,t) + c_j,
\eqno(53a)$$
where $b_j(y,t)$ is an arbitrary function of (y,t) satisfying the relation

$$b_j(y,t) = b_j(\rho) = b_j(y+ia_jt),\eqno(53b)$$
and $a_j$ and $c_j$ are complex constants. In order to construct the one
soliton solution, we take the case of $N = 1$ in (53a) and substitute it in
Eq. (51a). We obtain

$$f_2=\exp {(\chi_1+\chi_1^*+\psi)}.\eqno(54a)$$
Using Eqs. (51a) and (52a), it is found that

$$\exp {\psi} = -{a_1^2 \over {(a_1+a_1^*)^2}}. \eqno(54b)$$
If we use the above forms of $g_1$ and $f_2$ in Eqs. (50b), (51b) and (52b), 
we can see that $g_j=0$ for $j\ge 3$ and $f_j=0$ for $j\ge 4$. By substituting the values
of $g_1$ and $f_2$ in Eqs. (48a) and (48b), we obtain the expressions for the
1-soliton solution for the spin components and for the potential for example 
with the choice $\exp c_1 = {2a_{1R}\over {a_1^*}}$ as 

$$S^+(x,y,t)={2a_{1R} \over {a_{1R}^2+a_{1I}^2}} \exp {i\chi_{1I}}\left[ ia_{1I}
-a_{1R}tanh \chi_{1R}\right] sech \chi_{1R}, \eqno(55a)$$

$$S_3(x,y,t)=1-{2 a_{1R}^2 \over {a_{1R}^2+a_{1I}^2}} sech^2{\chi_{1R}},
\eqno(55b)$$
and the potential as 
                       
$$u(x,y,t)={2a_{1R}\over {a_{1R}^2+a_{1I}^2}} \left(a_{1I}b_{1R}'-a_{1R}
b_{1I}'\right) sech^2{\chi_{1R}}.\eqno(55c)$$
It follows from Eqs. (55) that one can identify two types of solitons:
\subsubsection{Line solitons:}

If we choose the function $b_1(y,t)$ in Eq. (53b) as 
$$b_1(y,t) = b_1y+ib_1a_1t,\eqno(56a)$$
then 
$$\chi_1 = a_1x+b_1y+ia_1b_1t+c_1\eqno(56b)$$
where $b_1$ is now a complex constant, and Eqs. (55) correspond to the line solitons. 
In this case the spin vector $\vec S\rightarrow (0,0,1)$ for fixed $t$ as $x$, $y
\rightarrow \pm \infty $, except along the line 
$$\chi_{1R} = a_{1R}x+b_{1R}y-(a_{1R}b_{1I}+a_{1I}b_{1R})t+c_{1R}=0,\eqno(57)$$
where it is still bounded.
\subsubsection{Curved solitons:}

However for arbitrary form of $b_1(\rho ) = b_1(y+ia_1t)$ in Eq. (53b) and for fixed
$(y,t)$, it follows from (55) that  $\vec S\rightarrow (0,0,1)$ as
$x,y \rightarrow \pm \infty$ and the wavefront itself is defined by the equation

$$\chi_{1R} = a_{1R}x + b_{1R}(\rho ) + c_{1R} = 0.\eqno(58)$$
We may call such solitons (which do not decay along the curve(58) ) as
curved solitons[23].
\subsection{The 2-soliton and N-soliton solutions}

To generate a 2 line or curved soliton solution (2-SS), we take $N=2$ in (53a) 
and hence $g_1$ takes the form

$$g_1=\exp{\chi_1}+\exp{\chi_2}.\eqno(59)$$
Substituting (59) in (50)-(52), after some calculation we obtain

$$f_2=N_{11}\exp {(\chi_1+\chi_1^*)}+N_{12}\exp {(\chi_1+\chi_2^*)}+N_{21}
\exp {(\chi_1^*+\chi_2)}+$$
$$N_{22}\exp {(\chi_2+\chi_2^*)},\eqno(60a)$$

$$g_3=L_{12}N_{11}N_{12} \exp {(\chi_1+\chi_1^*+\chi_2)}+L_{12}N_{22}N_{21}
\exp {(\chi_1+\chi_2+\chi_2^*)},\eqno(60b)$$

$$f_4=L_{12}L_{12}^*N_{11}N_{12}N_{21}N_{22} \exp {(\chi_1+\chi_1^*+\chi_2+
\chi_2^*)},\eqno(60c)$$
where

$$N_{rs}=-{a_r^2\over {(a_r+a_s^*)^2}},\,\,\,\,L_{rs}=-{(a_r-a_s)^2\over
{a_s^2}},\eqno(60d)$$
and $g_j=0$ for $j\ge 5$ and $f_j=0$ for $j\ge 6$. Inserting (60) into (48b)
we get

$$S^{+}(x,y,t) =2{(1+f^*_2+f^*_4)(g_1+g_3)\over {\mid 1+f_2+f_4\mid^2+\mid
g_1+g_3\mid ^2}},\eqno(61a)$$

$$S_{3}(x,y,t)={{\mid 1+f_2+f_4\mid^2-\mid g_1+g_3\mid ^2} \over {\mid 1+
f_2+f_4\mid^2+\mid g_1+g_3\mid ^2}}, \eqno(61b) $$
and similarly the expression for the potential can also be obtained from (48a)
or (46d).

Finally by taking $g_1$ as

$$g_1=\sum_{j=1}^N\exp{\chi_j}$$
and extending the above procedure, one can obtain the N-SS also.
\subsection{The domain wall type solution}

The soliton solutions of (55) and (61) correspond to the boundary condition

$$\vec S(x,y,t) = (0. 0, 1),\,\, as\,\,\, x,\,\,y \rightarrow \pm \infty .
\eqno(62)$$
Another class of physically interesting solutions are the domain wall
type solutions, which have the asymptotic form

$$\vec S(x,y,t) = (0,0,\pm 1), as\,\,\, x,\,\,y \rightarrow \pm \infty \eqno(63)$$
In order to obtain domain wall solutions in the present model, we make
the choice

$$\omega(x,y,t) = g(x,y,t), \,\,\,\,\, f(x,y,t) = 1 .  \eqno(64)$$
Then, Eq. (46) reduces to

$$ig_{t} + g_{xy} = 0,                 \eqno(65a)$$

$$g^{*}_{x} g_{y} + g^{*}_{y} g_{x} = 0,      \eqno(65b)$$

$$g^{*}_{x}g - g^{*}g_{x} = 0,                      \eqno(65c)$$
which is consistent with Eq. (13). Alternately, we can use another substitution

$$\omega (x,y,t) = \frac{1}{f(x,y,t)},  \,\,\,\,g(x,y,t) = 1 \eqno(66)$$
Here also it follows from (46) that

$$if^{*}_t - f^{*}_{xy} = 0     \eqno(67a)$$

$$f^{*}_{x}f_{y} +  f_{x}f^{*}_{y} = 0 \eqno(67b)$$

$$f^{*}_{x}f -  f^*f_{x} = 0. \eqno(67c)$$
Comparing Eqs. (65) and (67), we see that if $\omega (x,y,t)$ is a solution 
of Eq. (13), so also

$$ \omega ^{\prime }(x,y,t) =\pm {1\over {\omega (x,y,t)}} \eqno(68)$$
are solutions of Eq. (13). This is an obvious consequence of the fact that Eq. (13)
is invariant under inversion.

Now, we find the simplest non-trivial solutions for example of Eq. (65). Let us
take the ansatz

$$g=\exp {(ax+iby-abt)} \eqno (69)$$
where $a$, $b$ are real constants. The components of the spin vector $\vec S$ are
given by

$$S^+(x,y,t)={\exp {iby}\over {cosh [a(x-bt-x_0)]}},\eqno(70a) $$

$$S_3(x,y,t)=-tanh [a(x-bt-x_0)].\eqno(70b) $$
We can also have a more general solution of the form

$$g=\exp {[ax+im(y,t)]},\eqno(71) $$
where $a$ is a real constant and $m(y,t)$ is an arbitrary function of $y$ and
$t$. From Eq. (65a), it follows that

$$m=m(\rho )=m(y+iat), \rho=y+iat. \eqno(72) $$
Expressions for the spin components are then given by

$$S^+(x,y,t)={\exp [iRe (m(\rho ))]\over {cosh [ax-Im (m(\rho ))]}},\eqno(73a) $$

$$S_3(x,y,t)=-tanh [ax+Re (m(\rho ))],\eqno(73b) $$
and the potential is

$$u(x,y,t)=-2m^{\prime }_R \{ 1+\exp {[-2(ax-m_I)]}\}^{-1},\eqno(74) $$
where the $\prime $ denotes the differentiation with respect to the real part
of the argument. Naturally even more general solutions can be obtained by taking
more general forms for $g(x,y,t)$ than (69) or (71).
\subsection{The breaking soliton solution}.

We have already noted in Sec.II that for the present system (4), we have a
non-isospectral problem, as the spectral parameter $\lambda $ satisfies Eq. (9).
The above presented solutions all correspond to the constant solution of
Eq. (9), namely $\lambda =\lambda_1=$ constant. One may consider other
interesting solutions of Eq. (9). For example, one can have a special solution

$$\lambda=\lambda_1= \delta(y,t)+i \xi (y,t)={y+k+i\eta \over {q-t}},\eqno(75)$$
where $q$, $k$ and $\eta$ are real constants. Corresponding to this case, we may call
the resulting solutions of Eqs. (4) and (24) as breaking solitons[4]. Using 
the Hirota method, one can also construct the breaking 1-SS of Eq. (4) associated 
with (75). For this purpose, we take $g_1$ in the form

$$g=g_1= \exp{\chi },\,\,\,\, \chi = ax+m+c =\chi_R+i\chi_I ,\eqno(76) $$
where $a=a(y,t)$, $m=m(y,t)$ and $c=c(t)$ are functions to be determined.
Substituting (76) into the first of Eq. (53), we get

$$ia_t+aa_y=0,\,\,\,\,im_t+am_y=0,\,\,\,iA_t+Aa_y=0, \eqno(77)$$
where $A=\exp(c)$. Particular solutions of Eqs. (77) have the forms

$$ a=-i\lambda = {\eta -i(y+k) \over {q-t}},\,\,\,\, m=m{\left( y+k+i\eta
\over {{q-t}} \right) },\,\,\,\, A={A_0\over {q-t}} , \eqno(78)$$
where $\eta $, $k$, $q$ and $A_0$ are some constants. From Eqs. (50)-(52),
we obtain

$$f_2 =B \exp {2\chi_R},\,\,\,\, B={(y+k+i\eta )^2 \over
{4{\eta }^2}} .\eqno(79) $$

Now, we can write the breaking 1-SS of Eq. (4) (using equations (48b),
(76)-(79)),

$$ S^+(x,y,t) = {\exp{[i\chi_{1I}+\ln{2\eta \over {y+k+i\eta }}]} sech [\chi_{1R}+\ln
{y+k+i\eta \over {2\eta }}] \over {1+{\eta^2\over {(y+k)^2+\eta^2}} sech [\chi_{1R}
+\ln {y+k+i\eta \over {2\eta }}] sech [\chi_{1R}+\ln {y+k-i\eta\over {2\eta }}]}}
,\eqno(80a) $$

$$ S_3(x,y,t)= {1-{\eta^2 \over {(y+k)^2+\eta^2}} sech [\chi_{1R}+\ln {y+k+i\eta 
\over {2\eta }}] sech [\chi_{1R}+\ln {y+k-i\eta \over {2\eta }}]\over {1+
{\eta^2 \over {(y+k)^2+\eta^2}} sech [\chi_{1R}+\ln {y+k+i\eta 
\over {2\eta }}] sech [\chi_{1R}+\ln {y+k-i\eta \over {2\eta }}]}}
, \eqno(80b)$$
where 
$ \chi_1= \chi$ as defined in Eq. (76).
We see that the solution (80) corresponds to an algebraically decaying solution
for large $x$, $y$.
\subsection{Localized coherent structures (dromions)}

Next, we present the dromion type localized solutions of Eq. (4), the so-called
induced localized structures/or induced dromions[23] for the potential $u(x,y,t)$.
This is possible by utilising the freedom in the choice of the arbitrary
functions $b_{1R}$ and $b_{1I}$ of $b_1$ in Eq. (55c) and (53b).
For example, if we make the ansatz

$$b_{1I}(\rho_R)= k b_{1R}(\rho_R)= tanh(\rho_R), \eqno (81)$$

$$u=2 \eta (\xi -\eta k) sech^2{\rho_R}sech[\eta x+ tanh{\rho_R}-
\eta x_0], \eqno (82)$$
where $\rho _R=y-a_{1I}t$ and $k$ is a constant. Similarly, the expressions
for the spin can be obtained from Eqs. (55 a,b). The solution (82) for $u(x,y,t)$
decays exponentially in all the directions, eventhough the spin $\vec S $
itself is not fully localized. Analogously we can construct another type of
``induced dromion" solution with the choice

$$b_{1I}= k b_{1R} = \int {d \rho_R \over {(\rho_R+\rho_0)^2+1}}+b_0 ,
\eqno (83)  $$
where $\rho_0$ and $b_0$ are constants, so that

$$u(x,y,t)={2\eta (\xi - k y)\over {(\rho_R+\rho_0)^2+1}} sech^2 \left [\eta x+
\int {d \rho_R \over {(\rho_R+\rho_0)^2+1}}- \eta x_0\right].\eqno (84)$$
Proceeding in this way we can construct even more general solutions and multidromions
for the potential.
\section{SOLUTIONS OF (2+1) DIMENSIONAL NLSE}

In this section, we wish to consider briefly the corresponding solutions of the
equivalent generalized NLSE Eq. (24). Already this equation has received some
attention in the literature. The following types of solutions are available[23,24]:\\
a) Line solitons,\\
b) Induced dromions.\\
Now, we can construct the N-breaking soliton solutions of Eq. (24) as well. As 
an example, let us obtain the 1-breaking soliton solution of Eq. (24). The 
Hirota bilinear form of Eq. (24) can be obtained by using the transformation.

$$\psi={h\over \phi}\eqno(85)$$
as [20,24]

$$[iD_t+D_xD_y](h \circ \phi) = 0, \eqno(86a)$$

$$D_x^2(\phi\circ \phi)=2hh^*. \eqno(86b)$$
We look for the 1-breaking soliton solution in the following form:

$$h=\exp {\chi }, \eqno(87a)$$

$$\phi=1+\phi_2, \eqno(87b)$$
where$\chi =b(y,t)x+n(y,t)+c(t) $. Substituting (87) into (86), we get

$$ib_t+bb_y=0 , \eqno(88a)$$

$$in_t+bn_y=0 , \eqno(88b)$$

$$iB_t+Bb_y=0 , \eqno(88c)$$
and

$$\phi_2={1 \over {(b+b^*)^2}}\exp {\chi +\chi^*}=\exp {2(b_Rx+n_R
+\chi _0)}, \eqno (89) $$
where $\exp {2\chi_0}={1 \over {4b_R^2}}$, $B=\exp{c(t)}$ and
$b_R=b_R(t)=Re(b)$. Now, the formula (85) provides us the 1-breaking soliton
solution of Eq. (24),

$$\psi (x,y,t)={b_R(t)\exp {i\left[ b_I(y,t)x+n_I(y,t)+c_0\right] } \over
cosh \left[ b_Rx+n_R(y,t)+\chi_0\right] }, \eqno(90)$$
where $b(y,t)=b_R+ib_I$, $n(y,t)=n_R+in_I$ and $B(t)$ are the solutions of
Eqs. (88). Just as in the case of Eq. (77), if we take the following particular
solutions of the system of Eqs. (88);

$$b=-i\lambda ={\eta-i(y+k) \over {q-t}}, n={y+k+i\eta \over {q-t}},
B={B_0\over (q-t)},\eqno(91)$$
then the 1-breaking soliton solution of Eq. (24) takes the form

$$\psi (x,y,t)=-{\eta \over {q-t}}\exp{i\left[- {y+k\over {q-t}}x+n_I(y,t)
+c_0\right] }sech Z,\eqno(92)$$
where $Z={\eta \over {q-t}}x+n_R(y,t)+\chi_0$ and $c_0$, $\chi_0$ are constants.

Similarly, we obtain the breaking N-SS of (24). In this case we can take the
ansatz

$$g_{1} = \sum_{j = 1}^{N} \exp {\chi_{j}} \eqno(93)$$
with $\chi_{j} = b_{j}(y,t) x + n_{j}(y,t) + c_{j}(t)$. Inserting (93) into
(92), one is lead to

$$ib_{jt}+b_jb_{jy}=0,  \eqno(94a)$$

$$in_{jt}+b_jn_{jy}=0,  \eqno(94b) $$

$$iB_{jt}+B_jb_{jy}=0,  \eqno(94c) $$
Proceeding as before, one can obtain breaking N-soliton solution.
\section{SIMPLEST INTEGRABLE EXTENSIONS}

As mentioned above in ref.[5] (see also [25,26]) a new class of (2+1) dimensional
integrable spin equations was proposed. In particular, Eq. (4) is a particular
reduction of the following so called M-III equation (according
to the notations of ref.[5])

$$\vec S_{t} = (\vec S \wedge \vec S_{y} + u \vec S)_{x} +
2l (cl + d)\vec S_{y} - 4cV\vec S_{x} + \vec S\wedge A\vec S,\eqno(95a)$$

$$u_{x} = -\vec S\cdot (\vec S_{x} \wedge \vec S_{y}),  \eqno(95b)$$

$$V_{x} = \frac {1}{4(2lc+ d)^{2}} (\vec S^{2}_{x})_{y}, \eqno(95c)$$
where $l$, $c$ and $d$ are constants and $A$ is the anisotropy tensor.
This equation possesses some integrable reductions. For example, one obtains\\
a) the isotropic M - I equation (4), when c = 0, \,\,\  A = 0;\\
b) the isotropic M - II equation, when d = 0, \,\,\, A = 0;\\
c) the isotropic M - III equation, when $c \neq 0 \neq d,\,\,\, A = 0$;\\
d) the anisotropic M - I equation, when c = 0 \\
and so on. All these equations are integrable in the sense that each one of such
reduction has a Lax representation[5]. As an example, let us present the
associated linear problem for the isotropic M - III equation[5],

$$\psi_{1x} = U_{1}\psi_1 ,   \eqno(96a)$$

$$\psi_{1t} = V_{1}\psi_1 + (2c\lambda^{2} + 2d\lambda)\psi_{1y},  \eqno(96b)$$
with

$$U_{1} = [ic(\lambda^{2} - l^{2}) + id(\lambda - l)] S +
\frac{c(\lambda - l)}{2cl + d} S S_{x}, \eqno(97a)$$

$$V_{1} = [2c(\lambda^{2} - l^{2}) + 2d(\lambda - l)]B + \lambda^{2}F_{2}
+ \lambda F_{1} + F_{0},       \eqno(97b)$$
where

$$F_{2} = - 4ic^{2}VS,$$

$$F_{1} = - 4icdVS - \frac{4c^{2}V}{2cl + d} V SS_{x} -
\frac{ic}{2cl+d} S\{(SS_{x})_{y} - [SS_{x},B] \},  \eqno(98)$$

$$F_{0} = - lF_{1} - l^{2}F_{2}.$$
Here

$$B = \frac{1}{4}([S,S_{y}] + 2iuS), S=\vec S\cdot \vec \sigma$$
From these equations one can deduce the corresponding Lax representations
of the isotropic M -I and M - II equations by choosing c = 0 and d = 0, respectively.
We note that Eq. (95) for the isotropic M-III equation reduction case
is gauge[27] and L-equivalent[14] to the following equation

$$i\phi_{t} = \phi_{xy} - 4ic(V\phi)_{x} + 2 d^{2} V\phi, \eqno(99a)$$

$$V_{x} = (\mid \phi \mid^{2})_{y}.             \eqno(99b)$$
This equation has two integrable cases: a) if c=0, Eq. (99) reduces
to Eq. (24); b) if d=0, we get the Strachan equation[19]. The Lax
representations corresponding to (99) is obtained as follows.

$$\psi_{2x} = U_{2}\psi_{2},       \eqno(100a)$$

$$\psi_{2t} = V_{2}\psi_{2} +(2c\lambda^{2} + 2d\lambda)\psi_{2y}, \eqno(100b)$$
with

$$U_{2}=i[(c\lambda^{2} + d \lambda)\sigma_{3}+(2c\lambda +d)Q],\eqno(101a)$$

$$V_{2} = \lambda^{2} B_{2} + \lambda B_{1} + B_{0}.  \eqno(101b)$$
Here

$$B_{2} = -4ic^{2}V\sigma_{3},   $$

$$B_{1} = -4idcV\sigma_{3} + 2c\sigma_{3}Q_{y} - 8ic^{2}VQ,   \eqno(102)$$

$$B_{0} = \frac{d}{2c}B_{1} - \frac{d^{2}}{4c^{2}}B_{2},$$

$$Q=\pmatrix{
0 & {\phi} \cr
-{\phi^*} & 0
}.$$
\section{CONCLUSIONS}

In this paper we have obtained many interesting classes of exact solutions of
Eq. (4) and its equivalent counterpart Eq. (24), after estabilishing their L-
equivalence and gauge equivalence. The equivalence concept was also extended
to more general cases in Sec.IX. Another interesting class of
solutions are the periodic solutions which for Eq. (4) are given by

$$S_1(x,y,t)={CD-AB\over {AD-BC}}, S_2(x,y,t)=-i{CD+AB\over {AD-BC}},$$

$$S_3(x,y,t)={AD+BC\over {AD-BC}},  \eqno(103) $$
where

$$A=\theta (z-{r\over 2}  -{\delta \over 2}, \tau),\,\,\
B=\theta (z+{r\over 2}  -{\delta \over 2}, \tau), $$

$$C=\theta (z-{r\over 2}  +{\delta \over 2}, \tau),\,\,\
D=\theta (z+{r\over 2}  +{\delta \over 2}, \tau). \eqno(104)$$
Here, $\theta (z, \tau )$ is the Riemann $\theta $ functions and $r$,
$\delta $ are some constants. More detailed investigation of the periodic
solution(103) will be considered elsewhere.

Also we would like to note that the extended version of the L-equivalence
method which we used in section III works out for many other (2+1) dimensional
spin equations as well, for instance, to the Ishimori equation[14]. In the later
case, the unit vectors $\vec e_{k}$, $k=1,2,3,$ satisfy the Eqs. (14) and (18), while the
vector $\vec e_{1} \equiv \vec S $ satisfies the Ishimori equation

$$\vec e_{1t} = \vec e_{1} \wedge (\vec e_{1xx} + \epsilon^{2}\vec e_{1yy})
+ u_{x}\vec e_{1y} + u_{y}\vec e_{1x},  \eqno(105a)$$

$$u_{xx} - \epsilon^{2}u_{yy} =
-2\epsilon^{2}\vec e_{1} \cdot (\vec e_{1x} \wedge \vec e_{1y}). \eqno(105b)$$
Proceeding as in sec.III, we can obtain an evolution equation for the
curvature $\kappa $ and torsion $\tau $ in the form

$$\kappa_t=\omega_{3x}+\tau \omega_2,\eqno(106a)$$

$$\tau_t=\omega_{1x}-\kappa \omega_2,\eqno(106b)$$
where
$$\omega_1={\tau \omega_3-\omega_{2x}\over {\kappa }},\,\,\,\,\omega_2=u_x
\gamma_2-\kappa_x-\epsilon^2\gamma_{3y}+\gamma_1\gamma_2,$$

$$\omega_3=-\kappa \tau+\kappa u_y+u_x\gamma_3+\epsilon^2\gamma_{2y}-\gamma_1
\gamma_3,\,\,\,\,\gamma_2={(u_{xx}-\epsilon^2u_{yy})\over {2\epsilon^2\kappa}} $$
and the values of $\gamma_1$ and $\gamma_3$ can be obtained from the relations
(19a) and (19c) respectively.
Then the complex transformation

$$\phi = a \exp{ib},           \eqno(107)$$
where

$$a = {1\over 2} [\kappa^2 + \gamma_2^2+\gamma_3^2-2\kappa \gamma_2]^{1\over 2},
\eqno(108a)$$

$$b=\partial _x^{-1}(\tau -{u_y\over 2}+{\gamma_2\gamma_{3x}-\gamma_3\gamma_{2x}
-\kappa \gamma_{3x}\over {\kappa^2+\gamma_2^2+\gamma_3^2-2\kappa \gamma_2}})
\eqno(108b)$$
satisfies the Davey-Stewartson equation for $\epsilon^2=-1$,

$$i\phi_{t} = \phi_{yy} - \phi_{xx} +2\phi u, \eqno(109a)$$

$$u_{yy} + u_{xx} =(\mid \phi \mid)^2_{xx}-(\mid \phi \mid)^2_{yy},\eqno(109b)$$
where $u$ is a function of $x$, $y$ and $t$. Here $\kappa $ and $\tau $ have
the forms(17). Thus the geometrical equivalence between (105) and (109) can be estabilished
using the generalized transformation (107).

To summarize, in this paper, we have shown the equivalence of the (2+1)
dimensional spin equation (4) and the generalized (2+1) dimensional NLSE (24).
Besides we have found interesting classes of exact solutions for Eq. (4) and
NLSE(24). However, many questions remain open and deserve further investigation.
These include the existence of, for example of localized coherent structures (other
than the induced dromion-like solution presented here), the determination of the
Hamiltonian structure and the finding of possible physical applications of the
solutions obtained above. Another interesting problem is the classification of
solitons by the values of the topological charge. These questions are being 
pursued further.
\section*{ACKNOWLEDGEMENTS}

This work of M.L. forms part of a Department of Science
and Technology, Government of India sponsored research project. R.M. wishes
to thank Bharathidasan University for hospitality during his visits
to Tiruchirapalli. S.V. acknowledges the receipt of a Junior Research
Fellowship from the Council of Scientific and Industrial Research, India.
We would like to thank Prof. J. Zagrodzinski for useful discussion which lead us
to find the explicit forms of the periodical solution(92).
\end{document}